\documentclass[12pt,a4paper,dvips,epsfig,openany]{article}
\usepackage{a4p,rotating}
\usepackage{epsfig,subfigure}
\usepackage{cite,mcite}
\usepackage{graphicx}
\usepackage{physics,fleqn}
\usepackage{l3_title,ifthen,Lep}
\journalname{Phys. Lett. B}
\date{February 11, 1999}
\preprint{99-17}
\Lep{2}
\newlength{\capindent}
\newlength{\capwidth}
\newlength{\figwidth}
\newcommand{\icaption}[2][!*!,!]{\hspace*{\capindent}%
  \begin{minipage}{\capwidth}
    \ifthenelse{\equal{#1}{!*!,!}}%
      {\caption{#2}}%
      {\caption[#1]{#2}}
  \end{minipage}}
\input{l3at183.def}

\begin{document}
\begin{titlepage}
\title{{\LARGE 
       Measurement of Mass and Width \\
          of the W Boson at LEP      \\}}

\author{The L3 Collaboration}

%
%
\begin{abstract}
  
  We report on measurements of the mass and total decay width of the W
  boson with the L3 detector at LEP.  W-pair events produced in $\EE$
  interactions between $161~\GeV$ and $183~\GeV$ centre-of-mass energy
  are selected in a data sample corresponding to a total luminosity of
  76.7~pb$^{-1}$.  Combining all final states in W-pair production,
  the mass and total decay width of the W boson are determined to be
  $\MW=80.61\pm0.15~\GeV$ and $\GW=1.97\pm0.38~\GeV$, respectively.

\end{abstract}
%
%
\submitted
\end{titlepage}
%
%
\section{Introduction}

For the 1997 data taking period, the centre-of-mass energy,
$\sqrt{s}$, of the $\EE$ collider LEP at CERN was increased to
$183~\GeV$.  This energy is well above the kinematic threshold of
W-boson pair production, $\EE\!\rightarrow\mathrm{W^+W^-}$.

Analysis of W-pair production yields important knowledge about the
Standard Model of electroweak interactions~\cite{standard_model}
through the measurements of the mass, $\MW$, and the total decay
width, $\GW$, of the W boson~\cite{LEP2YRMW}.  These parameters were
initially measured at $\mathrm{p \bar p}$
colliders~\cite{MWPPBAR,GWPPBAR}.

First direct measurements of $\MW$ in $\EE$ collisions were derived
from total cross section
measurements~\cite{l3-111,l3-120,ALEPH-161-MW,DELPHI-161,OPAL-161-MW},
mainly at the kinematic threshold of the reaction
$\EE\!\rightarrow\mathrm{W^+W^-}$, $\sqrt{s}=161~\GeV$, where the
dependence of the W-pair cross section on the W-boson mass is largest.
At centre-of-mass energies well above the kinematic threshold, the
mass and also the total width of the W boson are determined by
analysing the invariant mass of the W-boson decay
products~\cite{l3-130,ALEPH-172-MW,DELPHI-172-XS+MW,OPAL-172-XS+MW}.

In this letter we report on an improved determination of the mass and
the total width of the W boson.  The analysis is based on the data
sample collected in the year 1997 at an average centre-of-mass energy
of $183~\GeV$, corresponding to an integrated luminosity of
55.5~pb$^{-1}$.  The invariant mass distributions of 588 W-pair events
selected at this energy are analysed to determine $\MW$ and $\GW$.
The results based on the 1997 data are combined with our previously
published measurements based on the 1996 data collected at
centre-of-mass energies of $161~\GeV$ and
$172~\GeV$~\cite{l3-111,l3-120,l3-130}.

\section{Analysis of Four-Fermion Production}

During the 1997 run the L3 detector~\cite{l3-01-new} collected
integrated luminosities of 4.04~pb$^{-1}$, 49.58~pb$^{-1}$ and
1.85~pb$^{-1}$ at centre-of-mass energies of $181.70~\GeV$,
$182.72~\GeV$ and $183.79~\GeV$, respectively, where these
centre-of-mass energies are known to $\pm0.05~\GeV$~\cite{LEPECAL97}.
These data samples are collectively referred to as $183~\GeV$ data in
the following.

The W boson decays into a quark-antiquark pair, such as
$\mathrm{W^-\!\rightarrow\bar u d~or~\bar c s}$, or a
lepton-antilepton pair, $\mathrm{W^-\!\rightarrow\ell^-\bar\nu_\ell}$
($\ell=\e,\mu,\tau$); in the following denoted as $qq$, $\ell\nu$ or
$ff$ in general for both W$^+$ and W$^-$ decays.  Four-fermion final
states expected in W-pair production are $\LNLNG$, $\QQLNG$, and
$\QQQQG$, where $(\gamma)$ indicates the possible presence of
radiative photons.

The following Monte Carlo event generators are used to simulate the
signal and background reactions: KORALW~\cite{KORALW} and
HERWIG~\cite{HERWIG} ($\EEWWFFFFG$); EXCALIBUR~\cite{EXCALIBUR}
($\EEFFFFG$); PYTHIA~\cite{PYTHIA} ($\EEQQG,\mathrm{ZZ}(\gamma)$);
KORALZ~\cite{KORALZ} ($\EEMMG,~\TTG$); BHAGENE3~\cite{BHAGENE},
BHWIDE~\cite{BHWIDE} and TEEGG~\cite{TEEGG} ($\EEEEG$),
DIAG36~\cite{DIAG36} and LEP4F~\cite{LEP4F} (leptonic two-photon
collisions); PHOJET~\cite{PHOJET} (hadronic two-photon collisions).
The response of the L3 detector is modelled with the
GEANT~\cite{xsigel} detector simulation program which includes the
effects of energy loss, multiple scattering and showering in the
detector material.

The selections of the four-fermion final states are described in
detail in References~\citen{l3-111,l3-120} and~\citen{l3-155} for the
data collected at $\sqrt{s}=161~\GeV$, $172~\GeV$ and $183~\GeV$.
These analyses reconstruct the visible fermions in the final state,
\ie, electrons, muons, $\tau$ jets corresponding to the visible $\tau$
decay products, and hadronic jets corresponding to quarks.  In order
to select a pure sample of $\QQQQ$ events, the cut of 0.67 on the
neural-network output described in the $\QQQQ$ cross-section analysis
is applied~\cite{l3-155}.  Kinematic constraints as discussed below
are then imposed to improve the resolution in the measured fermion
energies and angles and to determine those not measured.  The
invariant mass of the W boson is obtained from its decay products.

The mass and the width of the W boson are determined by comparing
samples of Monte Carlo events to the data.  A reweighting procedure is
applied to construct Monte Carlo samples corresponding to different
mass and width values.  Using this method, effects of selection and
resolution are automatically taken into account.

\section{Event Reconstruction imposing Kinematic Constraints}
\label{sec:reco-kf}

The final states $\QQEN$, $\QQMN$ and $\QQQQ$ contain at most one
primary unmeasured neutrino.  For each event a kinematic fit is
performed in order to determine energy, $E_f$, polar angle,
$\theta_f$, and azimuthal angle, $\phi_f$, for all four fermions, $f$,
in the final state.  The kinematic fit adjusts the measurements of
these quantities for the visible fermions according to their
experimental resolutions to satisfy the constraints imposed, thus
improving their resolution.

Four-momentum conservation and equal mass of the two W bosons are
imposed as constraints, allowing the determination of the momentum
vector of the unmeasured neutrino.  For the energy constraint, the
exact centre-of-mass energies as given in the previous section are
used. For hadronic jets, the velocity $\beta_f=|\vec p_f|/E_f$ of the
jet is fixed to its measured value as systematic effects cancel in the
ratio.  For $\QQEN$ and $\QQMN$ events, this yields a two-constraint
(2C) kinematic fit, whereas for $\QQQQ$ events it is a five-constraint
(5C) kinematic fit.

Events with badly reconstructed hadronic jets are rejected by
requiring that the probability of the kinematic fit exceeds 5\%.  The
kinematic fit mainly improves the energy resolution and less the
angular resolution.  The resolutions in average invariant mass,
$\Minv$, typically improve by a factor of four for $\QQEN$ and $\QQMN$
events and a factor of six for $\QQQQ$ events.

For $\QQTN$ events, the decay products of the leptonically decaying W
boson contain at least two unmeasured neutrinos in the final state.
Therefore only the hadronically decaying W boson is used in the
invariant mass reconstruction.  The energies of the two hadronic jets
are rescaled by a common factor so that the sum of their energies
equals half the centre-of-mass energy, thus imposing equal mass of the
two W bosons.  The rescaling improves the resolution in invariant mass
by nearly a factor of four.  Since invariant masses of W bosons in
$\LNLN$ events cannot be reconstructed as the decay of both W bosons
involves neutrinos, $\LNLN$ events are not used in the analysis for W
mass and width.

\section{Fitting Method for Mass and Width}
\label{sec:fit-method}

The fitting procedure uses the maximum likelihood method to extract
values and errors of the W-boson mass $\MW$, and the total width
$\GW$, denoted as $\Psi$ for short in the following.  In fits to
determine $\MW$ only, the Standard Model relation $\GW =
3\GF\MW^3/(2\sqrt{2}\pi)(1+2\aqcd/(3\pi))$~\cite{LEP2YRWW} is imposed.
Otherwise, $\MW$ and $\GW$ are treated as independent quantities.

The kinematic fit imposing the equal-mass constraint determines the
weighted average of the two invariant W masses in an event, $\Minv$,
which is considered in the fit for mass and width.  The total
likelihood is the product of the normalised differential cross
section, $L(\Minv,\Psi)$, evaluated for all data events. For a given
four-fermion final state $i$, one has:
\begin{eqnarray}
L_{i}(\Minv,\Psi) & = & 
\frac{1}{f_i(\Psi)\sigma_{i}(\Psi)+\sigma_{i}^{\BG}}
\left[f_i(\Psi)
      \frac{\d\sigma_{i}      (\Minv,\Psi)}{\d\Minv} +
      \frac{\d\sigma_{i}^{\BG}(\Minv)}     {\d\Minv}
\right]
\,,
\end{eqnarray}
where $\sigma_{i}$ and $\sigma_{i}^{\BG}$ are the accepted signal and
background cross sections and $f_i(\Psi)$ a factor calculated such
that the sum of accepted background and reweighted accepted signal
cross section coincides with the measured cross section.  This way
mass and width are determined from the shape of the invariant mass
distribution only.  The total and differential cross sections of the
accepted background are independent of the parameters $\Psi$ of
interest. They are taken from Monte Carlo simulations.

The total and differential signal cross sections depend on $\Psi$.
For values $\Psi_{\fit}$ varied during the fitting procedure, these
cross sections are determined by a reweighting procedure applied to
Monte Carlo events originally generated with parameter values
$\Psi_{\gen}$.  The event weights $R_{i}$ are given by the ratio:
\begin{eqnarray}
R_{i}(p_1,p_2,p_3,p_4,k_{\gamma},\Psi_{\fit},\Psi_{\gen}) & = & 
\frac
{\left|{\cal M}^{\mathrm{4F}}_{i}(p_1,p_2,p_3,p_4,k_{\gamma},\Psi_{\fit})\right|^2}
{\left|{\cal M}^{\mathrm{CC03}}_{i}(p_1,p_2,p_3,p_4,k_{\gamma},\Psi_{\gen})\right|^2}\,,
\end{eqnarray}
where ${\cal M}_{i}$ is the matrix element of the four-fermion final
state $i$. The matrix elements are calculated for the generated
four-vectors, $(p_1,p_2,p_3,p_4,k_{\gamma})$, of the four fermions and
any radiative photons in the event.  Since the Monte Carlo sample used
for reweighting is based on the three Feynman graphs in W-pair
production (CC03~\cite{CCNC,LEP2YRWW,LEP2YREG}), the matrix element in
the denominator is calculated using only CC03 graphs.  The matrix
element in the numerator is based on all tree-level graphs
contributing to the four-fermion final state $i$.  The calculation of
matrix elements is done with the EXCALIBUR~\cite{EXCALIBUR} event
generator.

The total accepted signal cross section for a given set of parameters
$\Psi_{\fit}$ is then:
\begin{eqnarray}
  \sigma_{i}(\Psi_{\fit}) & = & 
  \frac{\sigma^{\gen}_{i}}{N^{\gen}_{i}} \cdot
  \sum_{j}R_{i}(j,\Psi_{\fit},\Psi_{\gen})\,,
\end{eqnarray}
where $\sigma^{\gen}_{i}$ denotes the cross section corresponding to
the total Monte Carlo sample containing $N^{\gen}_{i}$ events. The sum
extends over all Monte Carlo events $j$ accepted by the event
selection.  

Based on the sample of reweighted events, two methods are used to
obtain the accepted differential signal cross section in reconstructed
invariant mass $\Minv$.  Both methods take detector and selection
effects as well as $\Psi$-dependent changes of efficiencies and
purities properly into account.

In the box method~\cite{BOXMETHOD}, the accepted differential cross
section is determined by averaging Monte Carlo events inside a
$\Minv$-bin centred around each data event.
The size of the bin considered is limited by the requirement of
including no more than 1000 Monte Carlo events, yielding bin sizes of
about $\pm35~\MeV$ at the peak of the invariant mass distribution.  In
addition, the bin size must not be larger than $\pm250~\MeV$ around
$\Minv$.

In the spline method, the continuous function describing the accepted
differential cross section is obtained by using a cubic spline to
smooth the binned distribution of reconstructed invariant masses.  At
the kinematic limit of $\sqrt{s}/2$ the value of the spline is fixed
to zero, while at the lower bound of $65~\GeV$ the value of the spline
is fixed to the average over a $2~\GeV$ interval.  The spline contains
25 knots in total.  Four knots are placed at each endpoint with the
remaining knots placed such that an equal number of Monte Carlo events
separates each knot.

Both methods yield identical results within 15\% of the statistical
error.  For the numerical results quoted in the following, the spline
method is used.

The fit procedure described above determines the parameters without
any bias as long as the Monte Carlo describes photon radiation and
detector effects such as resolution and acceptance functions
correctly.  By fitting large Monte Carlo samples, typically a hundred
times the data, the fitting procedure is tested to high accuracy.  The
fits reproduce well the values of the parameters of the large Monte
Carlo samples being fitted.  Also, the fit results do not depend on
the values of the parameters $\Psi_{\gen}$ of the Monte Carlo sample
subject to the reweighting procedure.

The reliability of the errors given by the fit is tested by fitting
for each final state several hundred small Monte Carlo samples, each
the size of the data samples.  The width of the distribution of the
fitted central values agrees well with the mean of the distribution of
the fitted errors.

\section{Mass and Total Width of the W Boson}
\label{sec:fit-mwgw}

Based on the data collected at $172~\GeV$ and at $183~\GeV$, the mass
of the W boson is determined for each of the final states $\QQEN$,
$\QQMN$, $\QQTN$ and $\QQQQ$ in separate maximum likelihood fits.  For
mass fits in the $\QQQQ$ channel, the pairing algorithm to assign jets
to W bosons used in the event selection~\cite{l3-120,l3-155} is
changed.  The pairing yielding the highest likelihood in the 5C
kinematic fit is chosen.  The fraction of correct pairings is reduced
to 60\% for the best combination and it is 25\% for the second best
combination.  However, the signal-to-background ratio in the relevant
signal region around $\Minv\approx80~\GeV$ is improved.  The loss of
correct pairings is recovered by including the pairing with the second
highest likelihood as an additional distribution.  Monte-Carlo studies
show that the two values for $\MW$ obtained from fitting separately
the distributions of the best and the second best pairing have a
correlation of $(-1.3\pm1.0)\%$, which is negligible.

\begin{figure}[p]
\begin{center}
{\epsfig{file=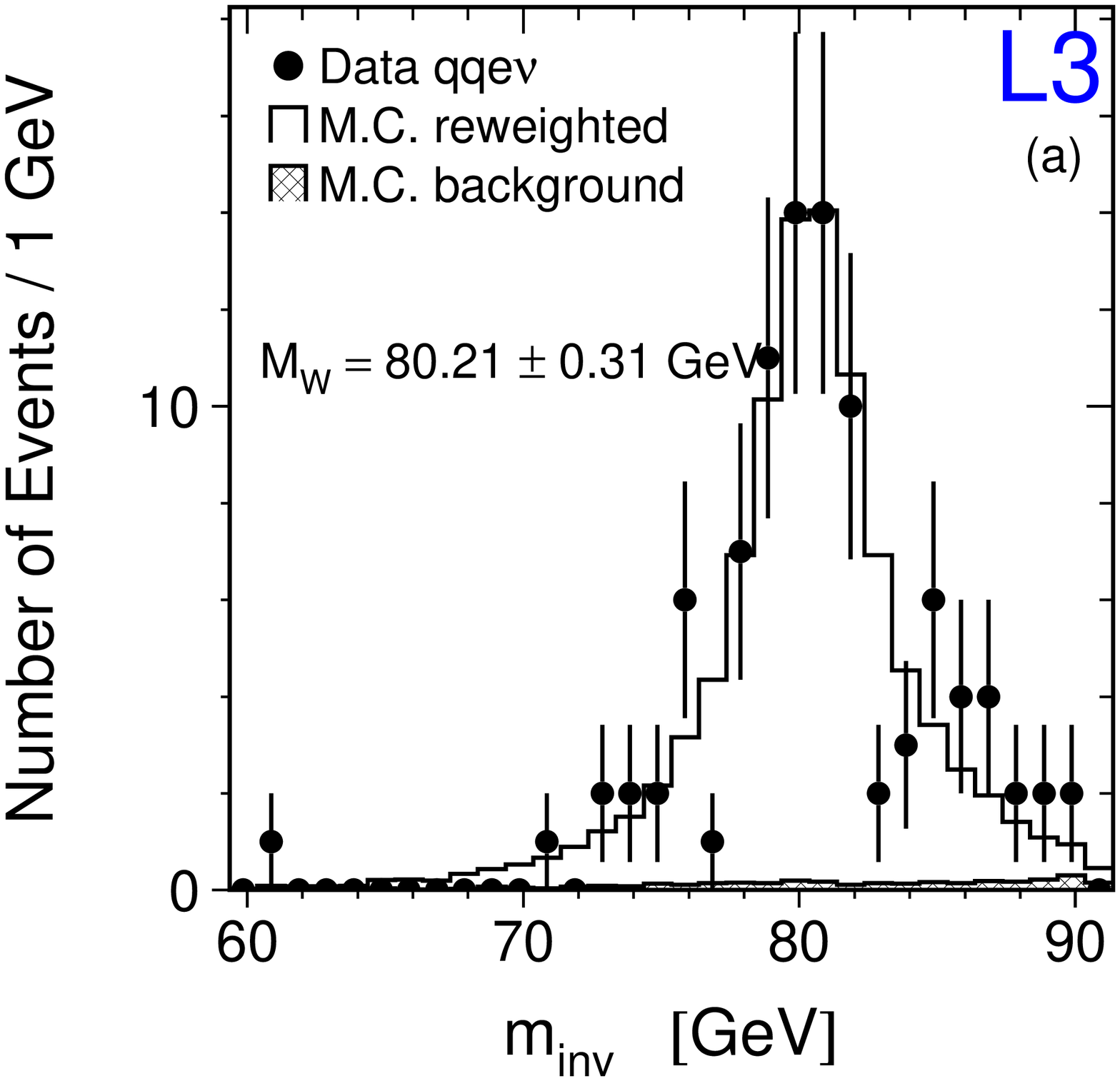,width=0.49\linewidth}}
{\epsfig{file=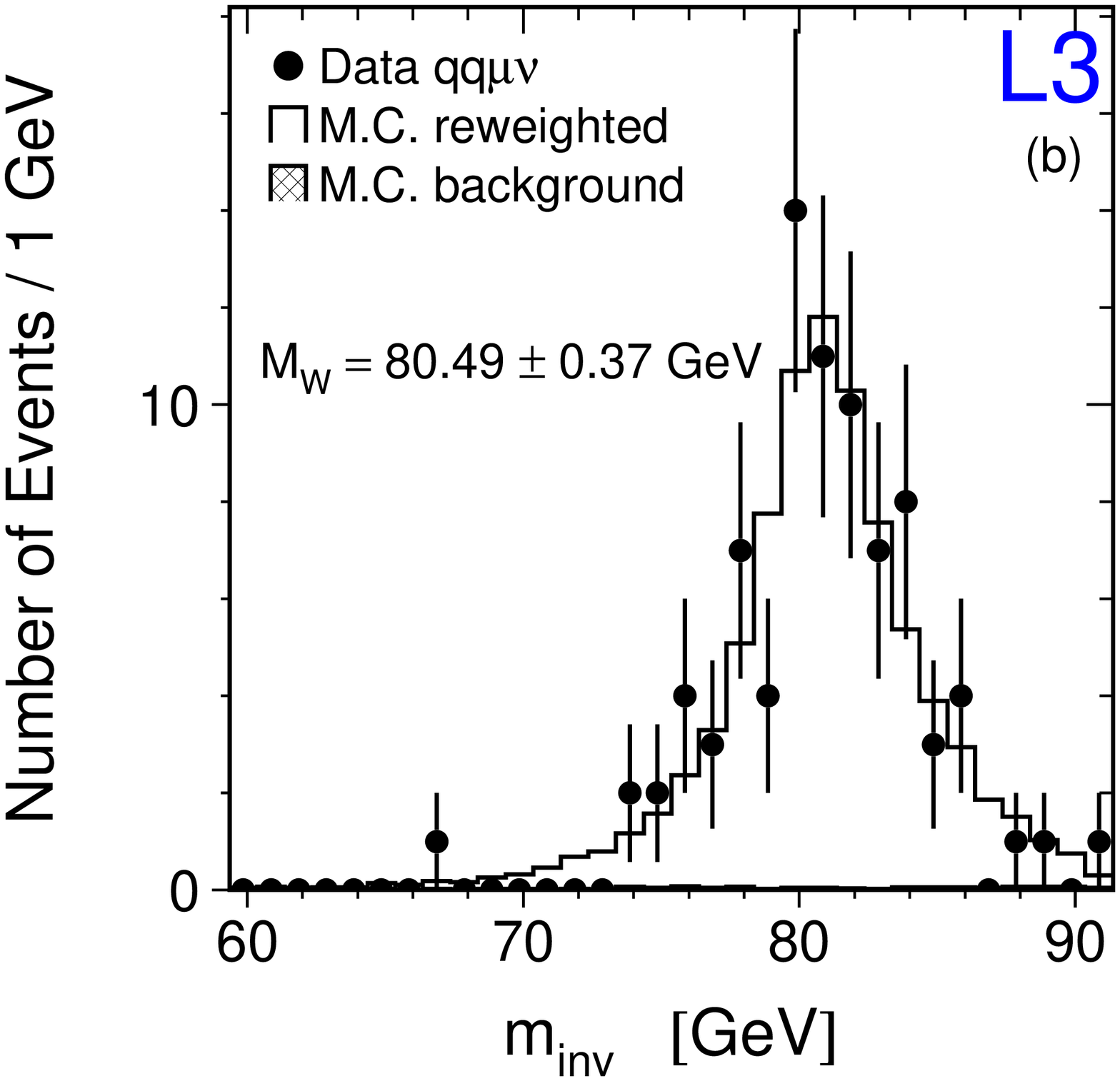,width=0.49\linewidth}}
\vskip -1.0cm
{\epsfig{file=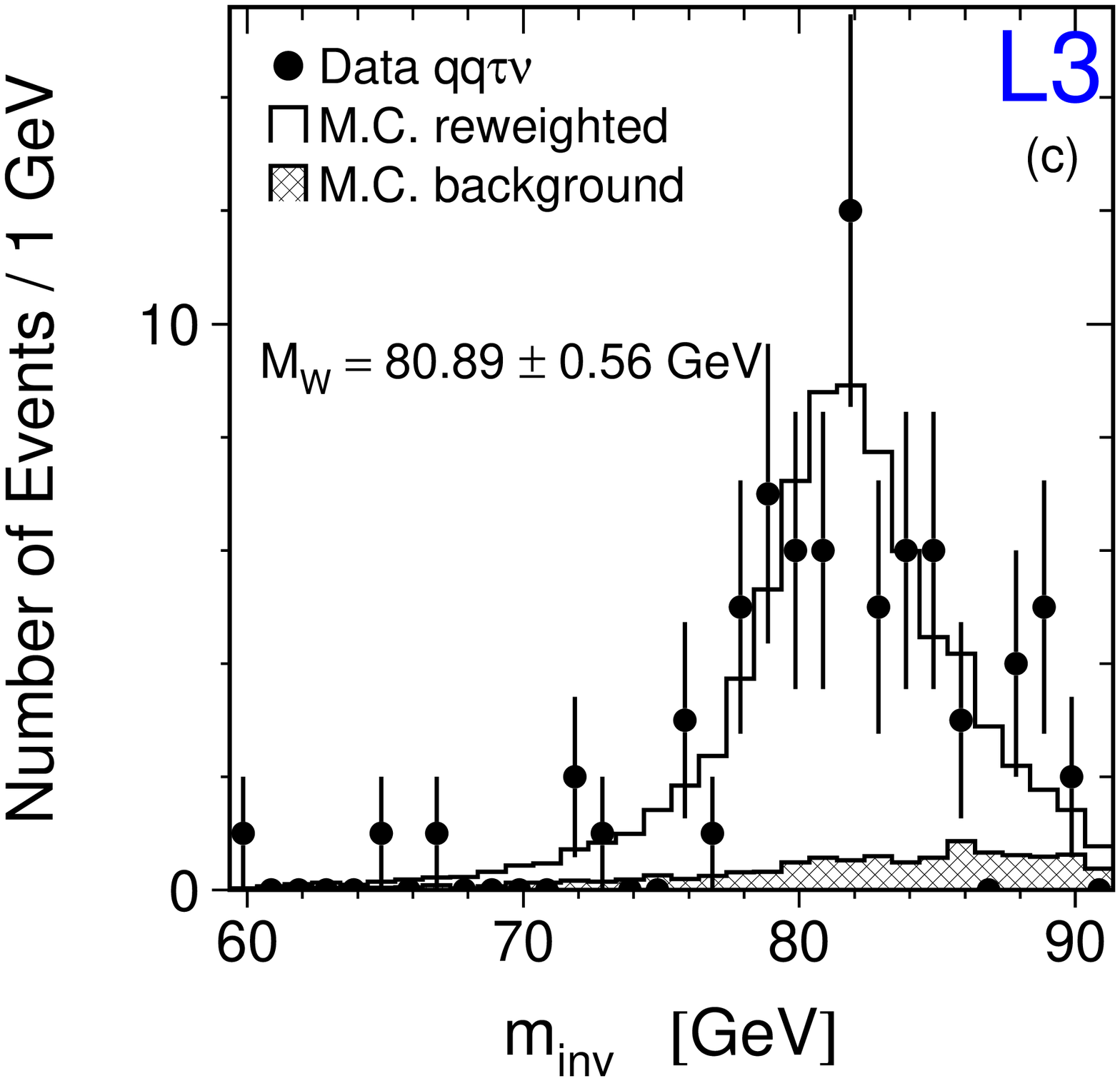,width=0.49\linewidth}}
{\epsfig{file=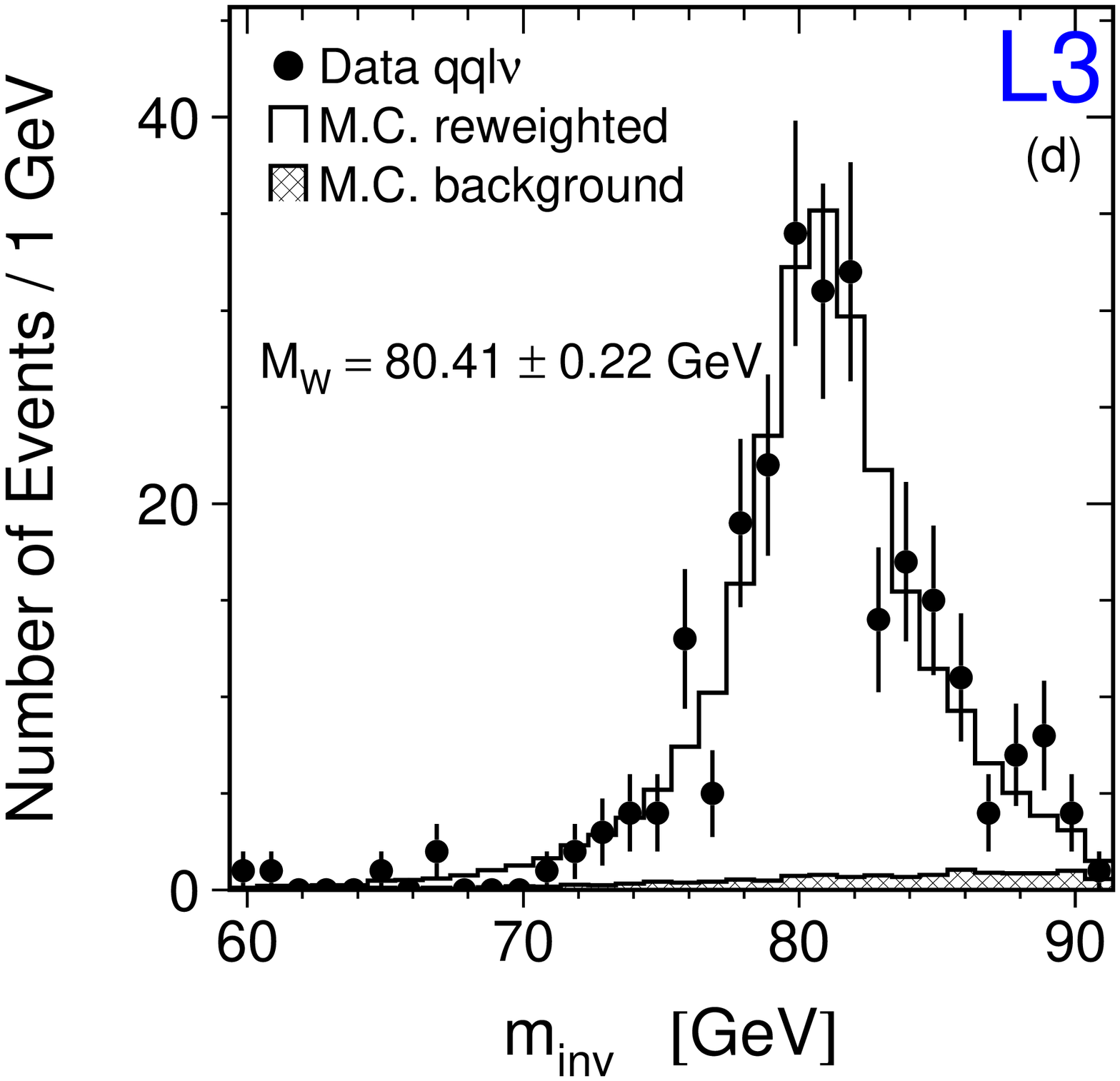,width=0.49\linewidth}}
\vskip -0.5cm
\caption[]{ 
  Distributions of reconstructed invariant mass, $\Minv$, after
  applying the kinematic fit using the equal-mass constraint for
  events selected in the $183~\GeV$ data: (a) $\QQEN$, (b) $\QQMN$,
  (c) $\QQTN$, (d) $\QQLN$, combining $\QQEN$, $\QQMN$ and $\QQTN$.
  The solid lines show the result of the fits of $\MW$ to the
  indicated final states. The quoted error combines statistical and
  systematic errors in quadrature. }
\label{fig:mw-minv-4}
\end{center}
\end{figure}

\begin{figure}[p]
\begin{center}
{\epsfig{file=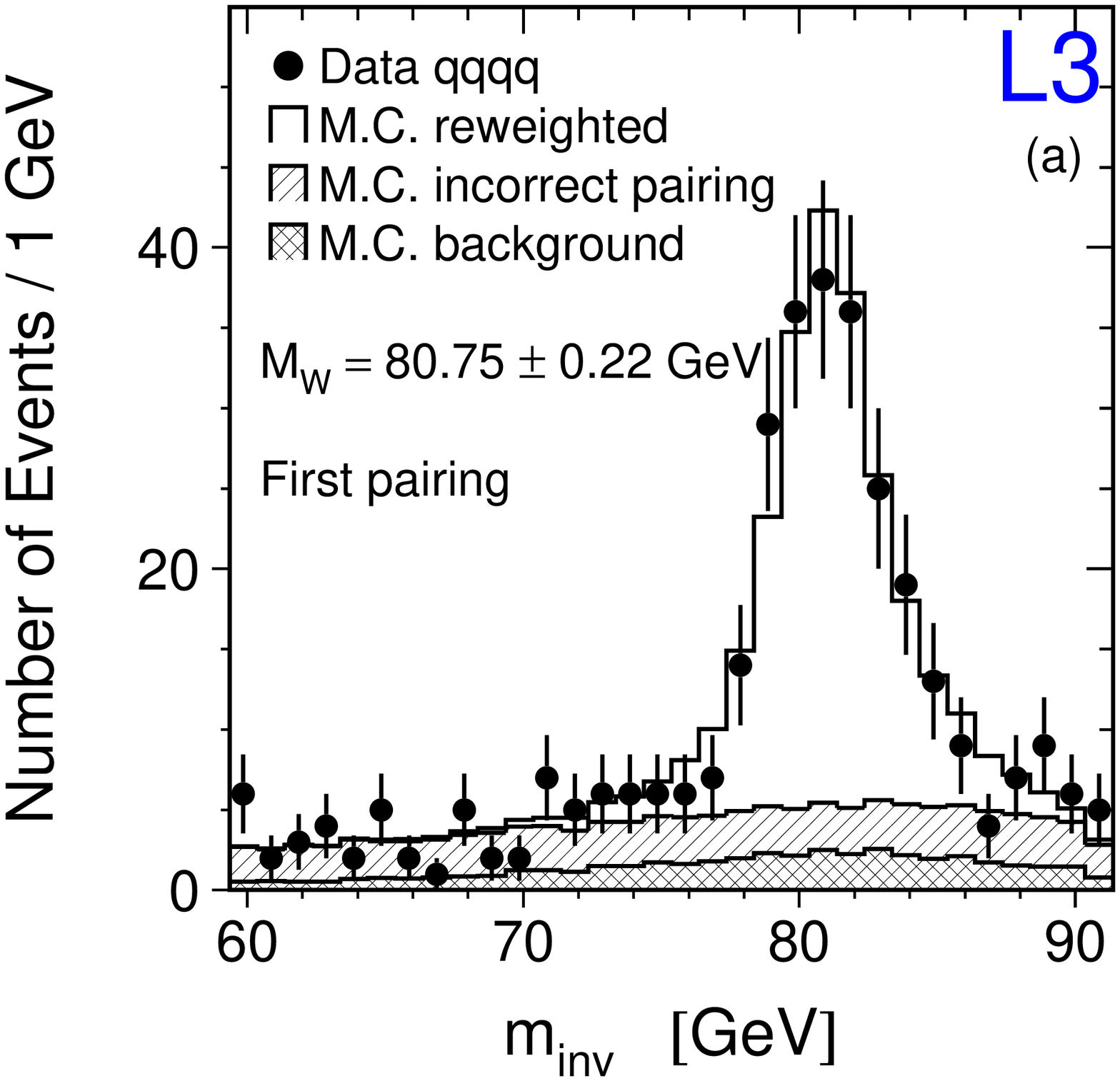,width=0.55\linewidth}}
\vskip -1.0cm
{\epsfig{file=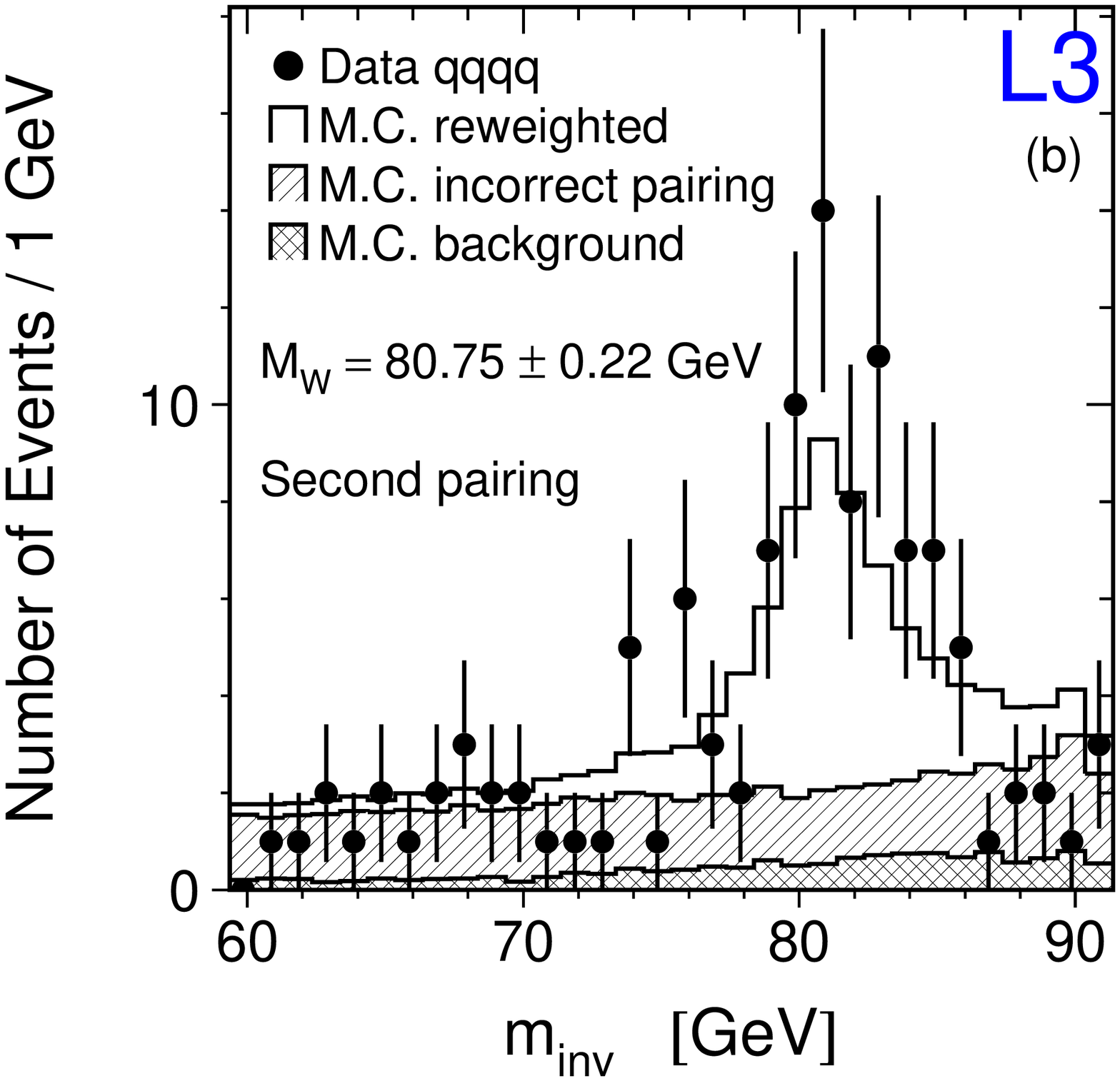,width=0.55\linewidth}}
\vskip -0.5cm
\caption[]{
  Distributions of reconstructed invariant mass, $\Minv$, after
  applying the kinematic fit using the equal-mass constraint for
  $\QQQQ$ events selected in the $183~\GeV$ data: (a) first pairing,
  (b) second pairing.  The solid lines show the result of the fit of
  $\MW$ to both pairings. The quoted error combines statistical and
  systematic errors in quadrature.}
\label{fig:mw-minv-2}
\end{center}
\end{figure}

\begin{figure}[p]
\begin{center}
{\epsfig{file=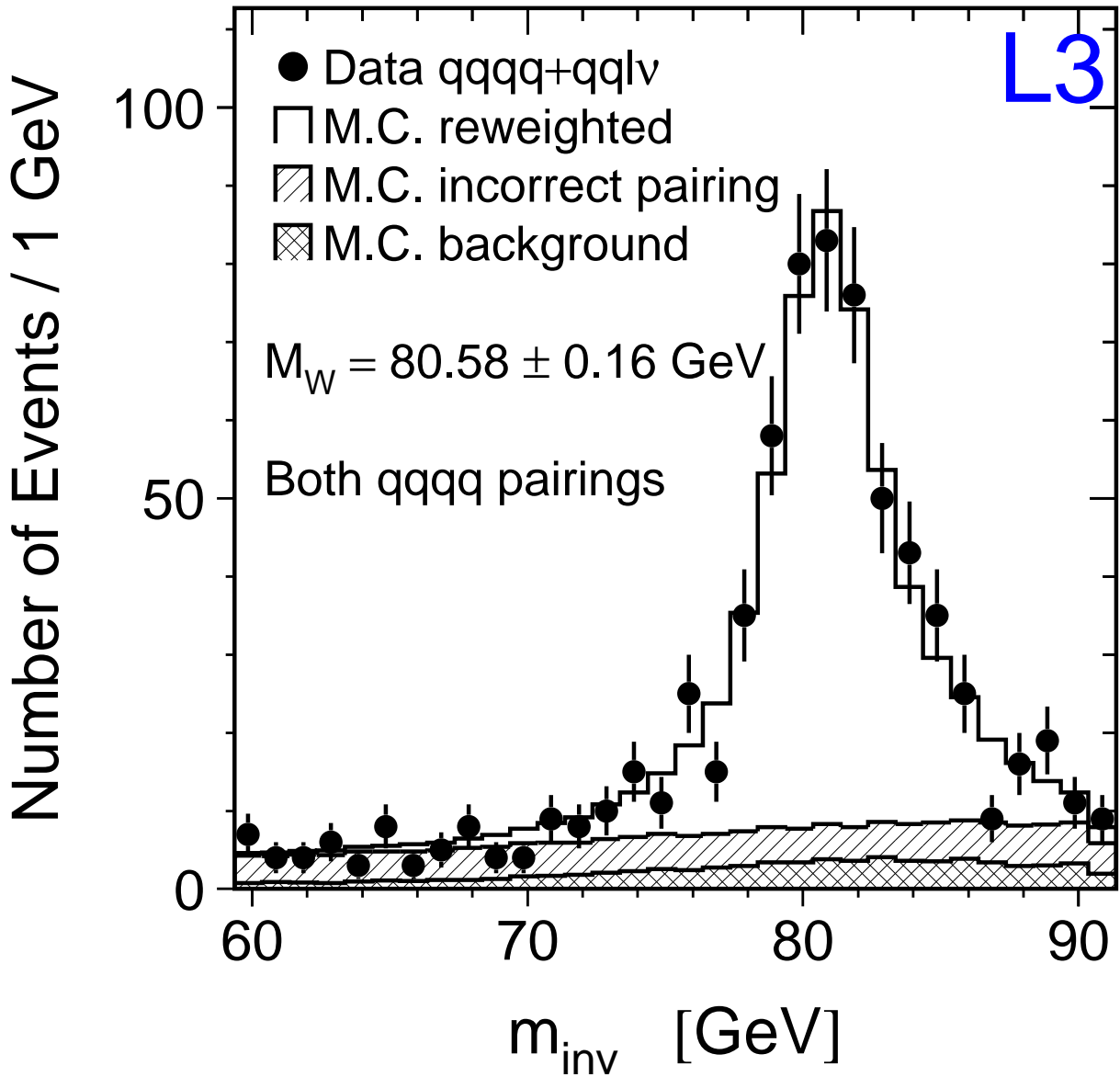,width=\linewidth}}
\caption[]{
  Distribution of reconstructed invariant mass, $\Minv$, after
  applying the kinematic fit using the equal-mass constraint for all
  W-pair events selected in the $183~\GeV$ data used for the mass
  analysis.  For $\QQQQ$ events, both pairings are included.  The
  solid line shows the result of the fit of $\MW$. The quoted error
  combines statistical and systematic errors in quadrature.}
\label{fig:mw-minv-1}
\end{center}
\end{figure}

The observed invariant mass distributions together with the fit
results for the semileptonic final states are shown in
Figure~\ref{fig:mw-minv-4}.  The distributions of the first and second
pairing in $\QQQQ$ events are shown in Figure~\ref{fig:mw-minv-2},
while the distribution summed over all final states and both $\QQQQ$
pairings is shown in Figure~\ref{fig:mw-minv-1}.  Combined results are
determined by averaging the results of individual channels taking
statistical and systematic errors into account.  The results of fits
for $\MW$ are summarised in Table~\ref{tab:mw-results-1d}.  The
observed statistical errors agree well with the statistical errors
expected for the size of the data samples used.  The results of fits
for $\MW$ and $\GW$ are summarised in Table~\ref{tab:mw-results-2d}.

\section{Systematic Effects}

The systematic errors on the fitted W mass and width are summarised in
Tables~\ref{tab:mw-syst} and~\ref{tab:gw-syst}.  They arise from
various sources and are divided into systematic errors correlated
between final states and systematic errors uncorrelated between final
states.

\subsection{Correlated Errors}

The beam energy of LEP is known with an accuracy of $25~\MeV$ for the
1997 data and $30~\MeV$ for the 1996 data, where $25~\MeV$ of these
errors are fully correlated~\cite{LEPECAL97}.  The relative error on
$\MW$ is given by the relative error on the LEP beam energy, while the
width is less affected.  The spread in centre-of-mass energy of about
$0.2~\GeV$ adds in quadrature to detector resolution and total width
of the W boson and is thus negligible.

Systematic uncertainties due to incomplete simulations of
initial-state radiation (ISR) are estimated by comparing the Monte
Carlo generators KORALW and EXCALIBUR implementing different QED
radiation schemes. For final-state radiation (FSR), events with FSR
simulation are compared to events without any FSR and a third of the
difference is taken as a systematic error.

The reconstruction of hadronic jets is examined by studying hadronic
$\QQG$ events collected at the Z pole and at $183~\GeV$.  A systematic
error for the jet measurement is assigned from varying the jet energy
scale by $0.2~\GeV$, smearing the jet energies by 5\% and smearing the
jet positions by $0.5^\circ$.  Effects due to fragmentation and
particle decays are determined by comparing signal events simulated
using string fragmentation as implemented in the PYTHIA Monte Carlo
program and cluster fragmentation as implemented in the HERWIG Monte
Carlo program to simulate the hadronisation process.

The fitting method itself is tested by fitting to various Monte Carlo
samples generated with known values for $\MW$ and $\GW$, varying over
a range of $\pm0.5~\GeV$.  The systematic error due to the fitting
method includes the effects due to different procedures for
reweighting and smoothing of the invariant mass distributions and
choice of technical parameters such as spline parameters, box size and
occupancy.

Limited Monte Carlo statistics introduces a tendency of the method to
have a slope of the linear function relating fitted mass to generated
mass less than one.  All Monte Carlo samples, approximately one
million events, are used in the reweighting procedure to minimise this
effect when fitting data.  Fitting several Monte Carlo samples and
using the remaining Monte Carlo as reference the non-linearity is
found to be negligible.


\subsection{Uncorrelated Errors}

The systematic error due to the size of the signal Monte Carlo sample
used for reweighting is estimated by dividing it into N parts of equal
size, N between 2 and 100, and making N fits to the same data sample.
The spread of the fit results, divided by the square root of N-1, is
found to be independent of N and yields the systematic error due to
Monte Carlo statistics.

Selection effects are estimated by varying the cut on the probability
of the kinematic fit and the interval of reconstructed invariant
masses being fitted.  Effects due to background are determined by
varying both the total accepted background cross section within its
error as evaluated for the cross section measurement as well as the
shape of the invariant mass spectrum arising from the background.

For $\QQQQ$ events, strong final state interactions (FSI) between the
hadronic systems of the two decaying W bosons due to effects of
colour-reconnection~\cite{FSI-CR,MC-CR-183-used} or Bose-Einstein
correlations~\cite{FSI-BE,MC-BEC-183-used} may affect the mass
reconstruction.  In both cases, possible effects are estimated by
comparing signal simulations including and excluding the modelling of
such effects and assigning the mass difference found as systematic
error.  In case of colour reconnection, two models, called
superconductor model type I and type II as implemented in PYTHIA~5.7
are studied~\cite{MC-CR-183-used}, adjusted such that they both yield
35\% reconnection probability.  In case of Bose-Einstein correlations,
the simulation of this effect as implemented in PYTHIA~5.7 is
used~\cite{MC-BEC-183-used}.

For $\QQEN$ and $\QQMN$ events, the reconstruction of the lepton
energy and angles also affects the invariant mass reconstruction.  In
analogy to hadronic jets, control samples of $\LLG$ events selected at
the Z pole are used to cross check the reconstruction of leptons.
Energy scales and resolutions are varied within their errors and the
resulting effect on W mass and width is quoted as a systematic error.

\subsection{Z Mass Reconstruction as Consistency Check}

All aspects of the mass measurement, ranging from detector calibration
and jet reconstruction to fitting method are checked using
$\EE\!\rightarrow\QQ\gamma$ events selected at $\sqrt{s}=183~\GeV$.
For such events, the hard initial-state radiative photon reduces the
centre-of-mass energy of the $\EE$ interaction.  The presence of the Z
resonance causes the distribution of the invariant mass of the jet-jet
system to exhibit a peak at the Z mass, as it originates from Z decay,
with a shape similar to the W mass spectrum.

A kinematic fit is used to improve the mass resolution, enforcing
four-momentum conservation in order to improve resolutions in energies
and angles of measured photons and of the two jets and to determine
the energy of one photon or two photons escaping along the beam axis.
For the extraction of the Z mass from the invariant mass spectrum the
same method as for the W mass measurement is applied.  Monte Carlo
events are reweighted according to the ratio:
\begin{eqnarray}
R_\mathrm{Z}(\sqrt{s^\prime},M^\mathrm{Z}_{\fit},M^\mathrm{Z}_{\gen})
& = & 
\frac{\frac{\d\sigma}{\d\sqrt{s^\prime}}(\sqrt{s^\prime},M^\mathrm{Z}_{\fit})}
{\frac{\d\sigma}{\d\sqrt{s^\prime}}(\sqrt{s^\prime},M^\mathrm{Z}_{\gen})}\,,
\end{eqnarray}
using the differential cross-section $\d\sigma/\d\sqrt{s^\prime}$
where $\sqrt{s^\prime}$ is the reduced centre-of-mass energy after
initial-state radiation at Monte Carlo generator level.

\begin{figure}[p]
\begin{center}
  {\epsfig{file=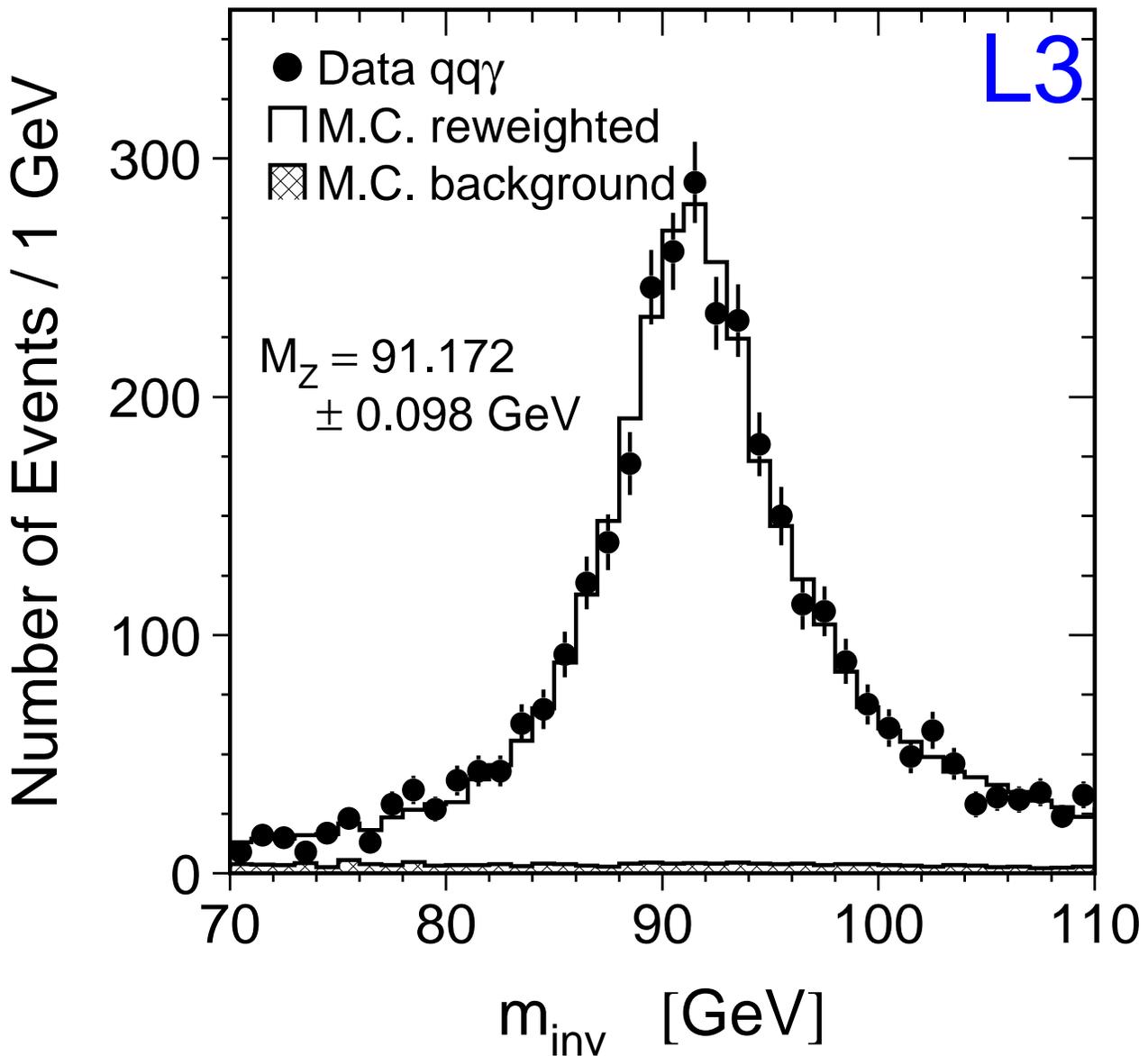,width=\linewidth}} 
\caption[]{
  Distribution of reconstructed invariant mass, $\Minv$, after
  applying the kinematic fit for $\QQ\gamma$ events with hard
  initial-state radiation selected at $183~\GeV$.  Shown is the region
  corresponding to the radiative return to the Z.  The solid line
  shows the result of the fit of $\MZ$. The quoted error is
  statistical. }
\label{fig:mz-minv-1}
\end{center}
\end{figure}

The reconstructed mass spectrum together with the fit result is shown
in Figure~\ref{fig:mz-minv-1}.  A total of 3351 events are selected in
a mass window ranging from $70~\GeV$ to $110~\GeV$.  The fitted Z-mass
value is $\MZ=91.172\pm0.098~\GeV$, where the error is statistical.
Within this error, the fitted Z mass agrees well with our measurement
of the Z mass derived from cross section measurements at
centre-of-mass energies close to the Z pole,
$\MZ=91.195\pm0.009~\GeV$~\cite{l3-069}.  The good agreement
represents an important test of the complete mass analysis method.

\section{Results}

The results on $\MW$ determined in the $\QQEN$, $\QQMN$, and $\QQTN$
final states are in good agreement with each other, as shown in
Table~\ref{tab:mw-results-1d}.  They are averaged taking statistical
and systematic errors including correlations into account, and
compared to the result on $\MW$ determined in the $\QQQQ$ final state,
also shown in Table~\ref{tab:mw-results-1d}.
The systematic error on the mass derived from $\QQQQ$ events contains
a contribution from possible strong FSI effects.  Within the
statistical accuracy of these measurements there is no significant
difference between $\MW$ as determined in $\QQLN$ and $\QQQQ$ events:
\begin{eqnarray}
\Delta\MW & = & \MW(\QQQQ)-\MW(\QQLN) ~ = ~ 
                 0.35\pm0.28~(stat.)\pm0.05~(syst.)~\GeV\,.
\end{eqnarray}
For the calculation of the systematic error on the mass difference,
the systematic errors due to strong FSI are not included.

Averaging the results on $\MW$ obtained from the $\QQLN$ and $\QQQQ$
event samples, including also FSI errors, yields:
\begin{eqnarray}
\MW        & = & 80.58\pm0.14~(stat.)\pm0.08~(syst.)~\GeV\,.
\end{eqnarray}
The summed mass distribution is shown in Figure~\ref{fig:mw-minv-1}
and compared to the expectation based on this W-mass value. The good
agreement between the data and the reweighted mass spectrum is
quantified by the $\chi^2$ value of 26 for 30 degrees of freedom which
corresponds to a probability of 66\%.  The mass values obtained in
fits which determine both $\MW$ and $\GW$ are the same as before
within $20~\MeV$ while the error on the mass is unchanged.

\begin{figure}[p]
\begin{center}
{\epsfig{file=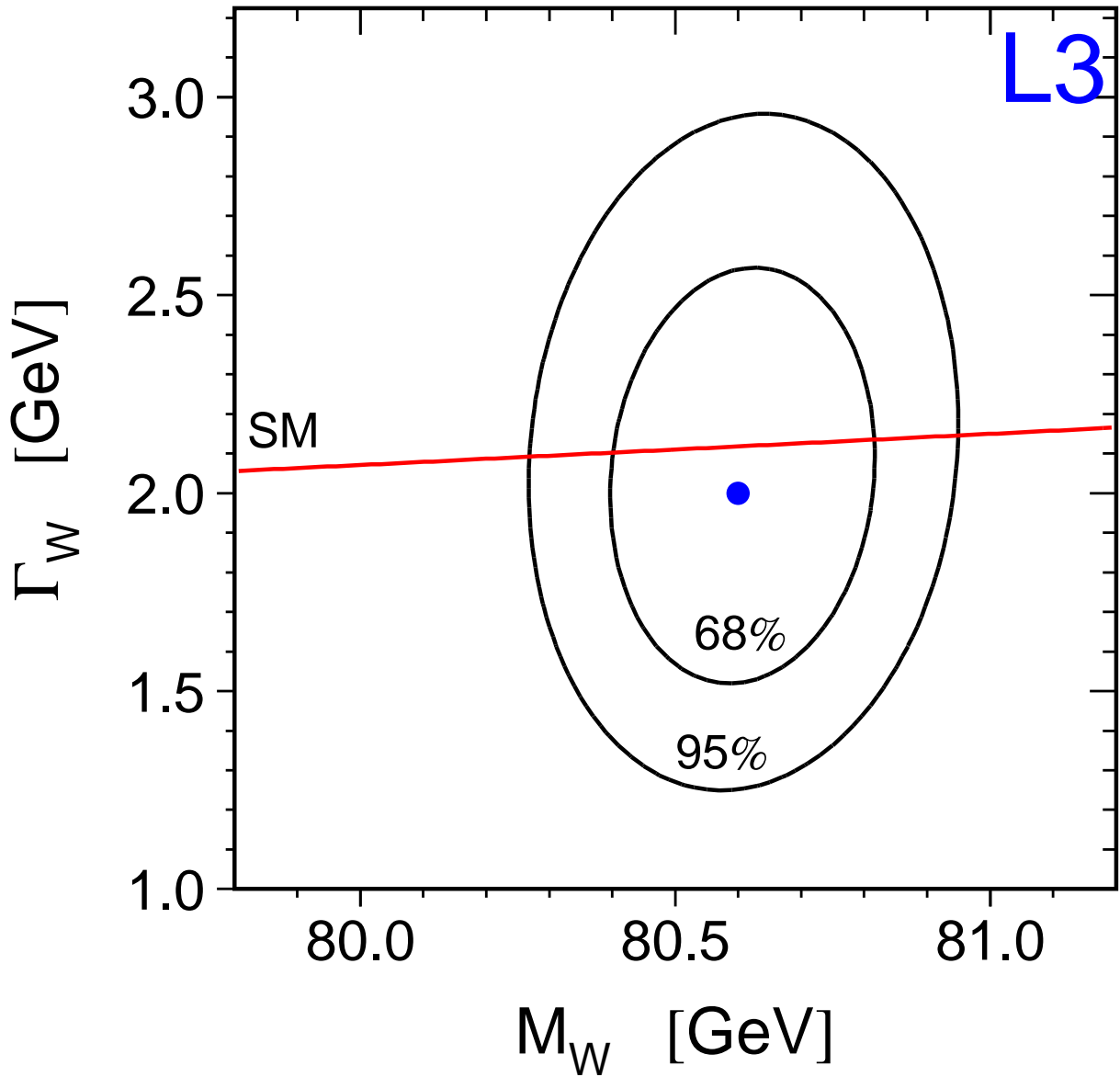,width=\linewidth}}
\caption[]{
  Contour curves of 68\% and 95\% probability in the $(\MW,\GW)$ plane
  from a fit to the combined $172~\GeV$ data and $183~\GeV$ data
  (statistical errors only).  The point represents the central values
  of the fit.  The Standard Model dependence of $\GW$ on $\MW$ is
  shown as the line. }
\label{fig:mw-gw}
\end{center}
\end{figure}

Within the statistical error, the width of the W boson determined in
$\QQQQ$ and $\QQLN$ events agree as shown in
Table~\ref{tab:mw-results-2d}.  For all final states combined the
result is:
\begin{eqnarray}
\GW        & = & 1.97\pm0.34~(stat.)\pm0.17~(syst.)~\GeV\,,
\end{eqnarray}
with a correlation coefficient of $+10\%$ between $\MW$ and $\GW$ as
shown in Figure~\ref{fig:mw-gw}.  Our result on $\GW$ is in good
agreement with the indirect measurement at $\mathrm{p \bar p}$
colliders, $2.07\pm0.06~\GeV$~\cite{GWPPBAR}, and measurements at
LEP~\cite{OPAL-172-XS+MW,OPAL-183-MW}.  It also agrees well with the
Standard Model expectation, $2.08~\GeV$, calculated for the current
world-average W mass~\cite{PDG98}.

The results on $\MW$ presented here agree well with our result derived
from the measurements of the total W-pair production cross section,
$\MW =
80.78^{+0.45}_{-0.41}~(exp.)\pm0.03~(\mathrm{LEP})~\GeV$~\cite{l3-120}.
Combining both results yields:
\begin{eqnarray}
  \MW & = & 80.61\pm0.15~\GeV\,.
\end{eqnarray}
This direct determination of $\MW$ is in good agreement with the
direct determination of $\MW$ at $\mathrm{p \bar p}$
colliders~\cite{MWPPBAR} and at LEP at lower centre-of-mass
energies~\cite{ALEPH-161-MW, ALEPH-172-MW, DELPHI-161,
  DELPHI-172-XS+MW, OPAL-161-MW, OPAL-172-XS+MW} and at
$183~\GeV$~\cite{OPAL-183-MW}.  It also agrees with our indirect
determination of $\MW$ at the Z peak,
$\MW=80.22\pm0.22~\GeV$~\cite{l3-069}, testing the Standard Model at
the level of its electroweak corrections.

%
%
\section{Acknowledgements}

We wish to congratulate the CERN accelerator divisions for the
successful upgrade of the LEP machine and to express our gratitude for
its good performance. We acknowledge with appreciation the effort of
the engineers, technicians and support staff who have participated in
the construction and maintenance of this experiment.

\clearpage

%
%
\bibliographystyle{l3stylem}

%
%
\newpage
\typeout{   }     
\typeout{Using author list for paper 171 -?}
\typeout{$Modified: Tue Feb  9 18:06:48 1999 by clare $}
\typeout{!!!!  This should only be used with document option a4p!!!!}
\typeout{   }
%
%
%
%
%
\input{Lep.sty}%
\ifx\LepCalled\undefined%
\typeout{     }%
\typeout{!!!!!!!!!!!!!!!!!!!!!!!!!!!!!!!!!!!!!!!!!!!!!!!!!!!!!!!!!!!}%
\typeout{Yikes.  You haven't used the Lep package!}%
\typeout{Please put \protect\usepackage\protect{Lep\protect} in your preamble,
         followed by}%
\typeout{\protect\Lep\protect{1\protect} or \protect\Lep\protect{2\protect}}%
\typeout{     }%
\typeout{For now you will get a Lep phase 2 authorlist (may not be right!).}%
\typeout{!!!!!!!!!!!!!!!!!!!!!!!!!!!!!!!!!!!!!!!!!!!!!!!!!!!!!!!!!!!}%
\typeout{     }%
\Lep{2}\fi%

\newcount\tutecount  \tutecount=0
\def\tutenum#1{\global\advance\tutecount by 1 \xdef#1{\the\tutecount}}
\def\tute#1{$^{#1}$}
\tutenum\aachen            
\tutenum\nikhef            
\tutenum\mich              
\tutenum\lapp              
\tutenum\basel             
\tutenum\lsu               
\tutenum\beijing           
\tutenum\berlin            
\tutenum\bologna           
\tutenum\tata              
\tutenum\ne                
\tutenum\bucharest         
\tutenum\budapest          
\tutenum\mit               
\tutenum\florence          
\tutenum\cern              
\tutenum\wl                
\tutenum\geneva            
\tutenum\hefei             
\tutenum\seft              
\tutenum\lausanne          
\tutenum\lecce             
\tutenum\losalamos         
\tutenum\lyon              
\tutenum\madrid            
\tutenum\milan             
\tutenum\moscow            
\tutenum\naples            
\tutenum\cyprus            
\tutenum\nymegen           
\tutenum\caltech           
\tutenum\perugia           
\tutenum\cmu               
\tutenum\prince            
\tutenum\rome              
\tutenum\peters            
\tutenum\salerno           
\tutenum\ucsd              
\tutenum\santiago          
\tutenum\sofia             
\tutenum\korea             
\tutenum\alabama           
\tutenum\utrecht           
\tutenum\purdue            
\tutenum\psinst            
\tutenum\zeuthen           
\tutenum\eth               
\tutenum\hamburg           
\tutenum\taiwan            
\tutenum\tsinghua          
{
\parskip=0pt
\noindent
{\bf The L3 Collaboration:}
\ifx\selectfont\undefined
 \baselineskip=10.8pt
 \baselineskip\baselinestretch\baselineskip
 \normalbaselineskip\baselineskip
 \ixpt
\else
 \fontsize{9}{10.8pt}\selectfont
\fi
\medskip
\tolerance=10000
\hbadness=5000
\raggedright
\hsize=162truemm\hoffset=0mm
\def\r{\rlap,}
\noindent

M.Acciarri\r\tute\milan\
P.Achard\r\tute\geneva\ 
O.Adriani\r\tute{\florence}\ 
M.Aguilar-Benitez\r\tute\madrid\ 
J.Alcaraz\r\tute\madrid\ 
G.Alemanni\r\tute\lausanne\
J.Allaby\r\tute\cern\
A.Aloisio\r\tute\naples\ 
M.G.Alviggi\r\tute\naples\
G.Ambrosi\r\tute\geneva\
% ...
\newpage
\begin{list}{A}{\itemsep=0pt plus 0pt minus 0pt\parsep=0pt plus 0pt minus 0pt
                \topsep=0pt plus 0pt minus 0pt}
\item[\aachen]
 I. Physikalisches Institut, RWTH, D-52056 Aachen, FRG$^{\S}$\\
 III. Physikalisches Institut, RWTH, D-52056 Aachen, FRG$^{\S}$
\item[\nikhef] National Institute for High Energy Physics, NIKHEF, 
     and University of Amsterdam, NL-1009 DB Amsterdam, The Netherlands
\item[\mich] University of Michigan, Ann Arbor, MI 48109, USA
\item[\lapp] Laboratoire d'Annecy-le-Vieux de Physique des Particules, 
     LAPP,IN2P3-CNRS, BP 110, F-74941 Annecy-le-Vieux CEDEX, France
\item[\basel] Institute of Physics, University of Basel, CH-4056 Basel,
     Switzerland
\item[\lsu] Louisiana State University, Baton Rouge, LA 70803, USA
\item[\beijing] Institute of High Energy Physics, IHEP, 
  100039 Beijing, China$^{\triangle}$ 
\item[\berlin] Humboldt University, D-10099 Berlin, FRG$^{\S}$
\item[\bologna] University of Bologna and INFN-Sezione di Bologna, 
     I-40126 Bologna, Italy
\item[\tata] Tata Institute of Fundamental Research, Bombay 400 005, India
\item[\ne] Northeastern University, Boston, MA 02115, USA
\item[\bucharest] Institute of Atomic Physics and University of Bucharest,
     R-76900 Bucharest, Romania
\item[\budapest] Central Research Institute for Physics of the 
     Hungarian Academy of Sciences, H-1525 Budapest 114, Hungary$^{\ddag}$
\item[\mit] Massachusetts Institute of Technology, Cambridge, MA 02139, USA
\item[\florence] INFN Sezione di Firenze and University of Florence, 
     I-50125 Florence, Italy
\item[\cern] European Laboratory for Particle Physics, CERN, 
     CH-1211 Geneva 23, Switzerland
\item[\wl] World Laboratory, FBLJA  Project, CH-1211 Geneva 23, Switzerland
\item[\geneva] University of Geneva, CH-1211 Geneva 4, Switzerland
\item[\hefei] Chinese University of Science and Technology, USTC,
      Hefei, Anhui 230 029, China$^{\triangle}$
\item[\seft] SEFT, Research Institute for High Energy Physics, P.O. Box 9,
      SF-00014 Helsinki, Finland
\item[\lausanne] University of Lausanne, CH-1015 Lausanne, Switzerland
\item[\lecce] INFN-Sezione di Lecce and Universit\'a Degli Studi di Lecce,
     I-73100 Lecce, Italy
\item[\losalamos] Los Alamos National Laboratory, Los Alamos, NM 87544, USA
\item[\lyon] Institut de Physique Nucl\'eaire de Lyon, 
     IN2P3-CNRS,Universit\'e Claude Bernard, 
     F-69622 Villeurbanne, France
\item[\madrid] Centro de Investigaciones Energ{\'e}ticas, 
     Medioambientales y Tecnolog{\'\i}cas, CIEMAT, E-28040 Madrid,
     Spain${\flat}$ 
\item[\milan] INFN-Sezione di Milano, I-20133 Milan, Italy
\item[\moscow] Institute of Theoretical and Experimental Physics, ITEP, 
     Moscow, Russia
\item[\naples] INFN-Sezione di Napoli and University of Naples, 
     I-80125 Naples, Italy
\item[\cyprus] Department of Natural Sciences, University of Cyprus,
     Nicosia, Cyprus
\item[\nymegen] University of Nijmegen and NIKHEF, 
     NL-6525 ED Nijmegen, The Netherlands
\item[\caltech] California Institute of Technology, Pasadena, CA 91125, USA
\item[\perugia] INFN-Sezione di Perugia and Universit\'a Degli 
     Studi di Perugia, I-06100 Perugia, Italy   
\item[\cmu] Carnegie Mellon University, Pittsburgh, PA 15213, USA
\item[\prince] Princeton University, Princeton, NJ 08544, USA
\item[\rome] INFN-Sezione di Roma and University of Rome, ``La Sapienza",
     I-00185 Rome, Italy
\item[\peters] Nuclear Physics Institute, St. Petersburg, Russia
\item[\salerno] University and INFN, Salerno, I-84100 Salerno, Italy
\item[\ucsd] University of California, San Diego, CA 92093, USA
\item[\santiago] Dept. de Fisica de Particulas Elementales, Univ. de Santiago,
     E-15706 Santiago de Compostela, Spain
\item[\sofia] Bulgarian Academy of Sciences, Central Lab.~of 
     Mechatronics and Instrumentation, BU-1113 Sofia, Bulgaria
\item[\korea] Center for High Energy Physics, Adv.~Inst.~of Sciences
     and Technology, 305-701 Taejon,~Republic~of~{Korea}
\item[\alabama] University of Alabama, Tuscaloosa, AL 35486, USA
\item[\utrecht] Utrecht University and NIKHEF, NL-3584 CB Utrecht, 
     The Netherlands
\item[\purdue] Purdue University, West Lafayette, IN 47907, USA
\item[\psinst] Paul Scherrer Institut, PSI, CH-5232 Villigen, Switzerland
\item[\zeuthen] DESY-Institut f\"ur Hochenergiephysik, D-15738 Zeuthen, 
     FRG
\item[\eth] Eidgen\"ossische Technische Hochschule, ETH Z\"urich,
     CH-8093 Z\"urich, Switzerland
\item[\hamburg] University of Hamburg, D-22761 Hamburg, FRG
\item[\taiwan] National Central University, Chung-Li, Taiwan, China
\item[\tsinghua] Department of Physics, National Tsing Hua University,
      Taiwan, China
\item[\S]  Supported by the German Bundesministerium 
        f\"ur Bildung, Wissenschaft, Forschung und Technologie
\item[\ddag] Supported by the Hungarian OTKA fund under contract
numbers T019181, F023259 and T024011.
\item[\P] Also supported by the Hungarian OTKA fund under contract
  numbers T22238 and T026178.
\item[$\flat$] Supported also by the Comisi\'on Interministerial de Ciencia y 
        Tecnolog{\'\i}a.
\item[$\sharp$] Also supported by CONICET and Universidad Nacional de La Plata,
        CC 67, 1900 La Plata, Argentina.
\item[$\natural$] Supported by Deutscher Akademischer Austauschdienst.
\item[$\diamondsuit$] Also supported by Panjab University, Chandigarh-160014, 
        India.
\item[$\triangle$] Supported by the National Natural Science
  Foundation of China.
\end{list}
}
\vfill



\newpage

\clearpage

\begin{table}[p]
\begin{center}
\renewcommand{\arraystretch}{1.3}
\begin{tabular}{|c||cc|c||c|}
\hline
Process &\multicolumn{2}{|c|}{Events} & Mass of the W Boson & Expected Stat. \\
\cline{2-5}
    & $172~\GeV$ & $183~\GeV$ &       $\MW~[\GeV]$   & Error $[\GeV]$   \\
\hline\hline
$\EEQQENG$ &   18 &   95 &       $80.21\pm0.30\pm0.06$  & $\pm0.31$ \\
$\EEQQMNG$ &    9 &   83 &       $80.49\pm0.36\pm0.06$  & $\pm0.34$ \\
$\EEQQTNG$ &   12 &   75 &       $80.89\pm0.56\pm0.08$  & $\pm0.47$ \\
\hline
$\EEQQLNG$ &   39 &  249 &       $80.41\pm0.21\pm0.06$  & $\pm0.21$ \\
\hline
$\EEQQQQG$ &   61 &  339 &       $80.75\pm0.18\pm0.12$  & $\pm0.20$ \\
\hline
$\EEFFFFG$ &   99 &  588 &       $80.58\pm0.14\pm0.08$  & $\pm0.14$ \\
\hline
\end{tabular}
\vskip 0.5cm
\caption[]{
  Number of events used in the analysis and results on the mass of the
  W boson, $\MW$, combining the data collected at $172~\GeV$ and at
  $183~\GeV$.  The first error is statistical and the second
  systematic.  Also shown is the statistical error expected for the
  size of the data sample analysed.  }
\label{tab:mw-results-1d}
\end{center}
\end{table}

\begin{table}[p]
\begin{center}
\renewcommand{\arraystretch}{1.3}
\begin{tabular}{|c||c|c|c|}
\hline
Process    & Mass of the W Boson & Total Decay Width & Correlation  \\
           &    $\MW~[\GeV]$     & $\GW~[\GeV]$      & Coefficient  \\
\hline\hline
$\EEQQLNG$ & $80.42\pm0.21\pm0.06$ & $2.44\pm0.59\pm0.13$ & $+0.10$ \\
$\EEQQQQG$ & $80.73\pm0.18\pm0.12$ & $1.69\pm0.42\pm0.22$ & $+0.15$ \\
\hline
$\EEFFFFG$ & $80.58\pm0.14\pm0.08$ & $1.97\pm0.34\pm0.17$ & $+0.10$ \\
\hline
\end{tabular}
\vskip 0.5cm
\caption[]{
  Results on the mass of the W boson, $\MW$, and its total decay
  width, $\GW$, combining the data collected at $172~\GeV$ and at
  $183~\GeV$.  The first error is statistical and the second
  systematic.  Also shown is the correlation coefficient between $\MW$
  and $\GW$. }
\label{tab:mw-results-2d}
\end{center}
\end{table}

\begin{table}[p]
\begin{center}
\renewcommand{\arraystretch}{1.2}
\begin{tabular}{|c||c|c|c|c|}
\hline
Systematic Errors  & \multicolumn{4}{|c|}{Final State}\\
\cline{2-5}     
 on $\MW~[\MeV]$   & ~$\QQEN$~&~$\QQMN$~&~$\QQTN$ &~$\QQQQ$ \\
\hline
\hline    
LEP Energy              &  25 &  25 &  25 &  25 \\ 
ISR                     &  15 &  15 &  15 &  15 \\ 
FSR                     &  10 &  10 &  10 &  10 \\ 
Jet Measurement         &  30 &  30 &  30 &   5 \\ 
Fragmentation and Decay &  30 &  30 &  30 &  60 \\ 
Fitting Method          &  15 &  15 &  15 &  15 \\ 
\hline
Total Correlated        &  55 &  55 &  55 &  69 \\
\hline
\hline
MC Statistics           &  20 &  20 &  50 &  10 \\ 
Colour Reconnection     & --- & --- & --- &  70 \\ 
Bose-Einstein Effects   & --- & --- & --- &  60 \\ 
Selection               &  20 &  20 &  20 &  20 \\ 
Background              &   5 &  10 &  30 &  10 \\ 
Lepton Measurement      &  15 &  15 & --- & --- \\ 
\hline
Total Uncorrelated      &  32 &  34 &  62 &  95 \\
\hline
\hline
Total Systematic        &  63 &  64 &  82 & 118 \\
\hline
\end{tabular}
\vskip 0.5cm
\caption[]{
  Systematic errors in the determination of $\MW$ for the different
  final states.  The contributions listed in the upper part of the
  table are treated as correlated when combining different final
  states.  The contributions listed in the lower part are treated as
  uncorrelated between channels. Total errors are obtained by adding
  the individual contributions in quadrature.}
\label{tab:mw-syst}
\end{center}
\end{table}

\begin{table}[p]
\begin{center}
\renewcommand{\arraystretch}{1.2}
\begin{tabular}{|c||c|c|}
\hline
Systematic Errors & \multicolumn{2}{|c|}{Final State}\\
\cline{2-3}
 on $\GW~[\MeV]$  &~$~\QQLN~$~&~$~\QQQQ~$~\\
\hline
\hline                                
LEP Energy              &  15 &  15 \\ 
ISR                     &  25 &  25 \\ 
FSR                     &  40 &  40 \\ 
Jet Measurement         &  80 &  20 \\ 
Fragmentation and Decay &  60 & 200 \\ 
Fitting Method          &  25 &  25 \\ 
\hline
Total Correlated        & 114 & 209 \\
\hline
\hline
MC Statistics           &  40 &  30 \\ 
Colour Reconnection     & --- &  50 \\ 
Bose-Einstein Effects   & --- &  10 \\ 
Selection               &  40 &  40 \\ 
Background              &  25 &  25 \\ 
Lepton Measurement      &  30 & --- \\ 
\hline
Total Uncorrelated      &  69 &  76 \\
\hline
\hline
Total Systematic        & 133 & 222 \\
\hline
\end{tabular}
\vskip 0.5cm
\caption[]{
  Systematic errors in the determination of $\GW$ in $\QQLN$ and
  $\QQQQ$ production.  The contributions listed in the upper part of
  the table are treated as correlated when combining the two final
  states.  The contributions listed in the lower part are treated as
  uncorrelated between channels. Total errors are obtained by adding
  the individual contributions in quadrature.}
\label{tab:gw-syst}
\end{center}
\end{table}

\clearpage



\end{document}